\documentclass[aps,reprint,showpacs,superscriptaddress,
footinbib,citeautoscript]{revtex4-1}

\usepackage{graphicx}

\usepackage[utf8]{inputenc}
\usepackage{csquotes}
\usepackage[american]{babel}
\usepackage[T1]{fontenc}
\usepackage{enumerate}
\usepackage{mdwlist}
\usepackage[activate=normal]{pdfcprot}
\usepackage{bbding}
\usepackage{color}
\usepackage{amssymb}
\usepackage{amsmath}
\usepackage{amsfonts}
\usepackage{mathrsfs}
\usepackage{lipsum}

\usepackage{bm}
\usepackage{dcolumn}
\usepackage{color}
\usepackage[colorlinks=true,citecolor=blue]{hyperref}
\hypersetup{colorlinks=true,citecolor=blue,linkcolor=red
,urlcolor=blue}

\begin{document}

\title{Thermoelectric transport in monolayer phosphorene}

\author{Moslem Zare}
\affiliation{School of Physics, Institute for Research in Fundamental Sciences (IPM), 19395-5531, Tehran, Iran}
\author{Babak Zare Rameshti}
\email{b.zare.r@ipm.ir}
\affiliation{School of Physics, Institute for Research in Fundamental Sciences (IPM), 19395-5531, Tehran, Iran}
\author{Farnood G. Ghamsari}
\affiliation{Department of Physics, Kharazmi University, 15719-14911, Tehran, Iran}
\affiliation{School of Physics, Institute for Research in Fundamental Sciences (IPM), 19395-5531, Tehran, Iran}
\author{Reza Asgari}
\affiliation{School of Physics, Institute for Research in Fundamental Sciences (IPM), 19395-5531, Tehran, Iran}
\affiliation{School of Nano Science, Institute
for Research in Fundamental Sciences (IPM), 19395-5531, Tehran,
Iran}

\begin{abstract}

We apply the generalized Boltzmann theory to describe thermoelectric transport properties of monolayer phosphorene in the presence of short- and long-range charged impurity interactions. First, we propose a low-energy Hamiltonian to explore the accurate electronic band structure of phosphorene in comparison with those results obtained by density-functional simulations. We explain the effect of the coupling between the conduction and valence bands on the thermoelectric properties. We show that the electric conductivity of phosphorene is highly anisotropic, while the Seebeck coefficient and figure of merit, without being influenced via either the presence or absence of the coupling term, are nearly isotropic. Furthermore, we demonstrate that the conductivity for the $n$ type of doping is more influenced by the coupling term than that of the $p$ type. Along with thermopower sign change, profound thermoelectric effects can be achieved.
\end{abstract}

\maketitle

\section{Introduction}\label{sec:intro}

Thermoelectric materials, based on a fundamental interplay between their electronic and thermal properties, have attracted much interest for application in energy conversion devices~\cite{Dresselhaus, Harman, Venkatasubramanian, Arita, Hamada, Zide, Wei, Zuev, Kato, Buscema, Konabe1}. The efficiency of the thermoelectric devices is quantified by a dimensionless figure of merit $\mathcal{Z}T$, which relates the Seebeck coefficient (thermopower) to the thermal conductivity. The small thermal conductivity and relatively high thermopower and electrical conductivity are required for high efficiency thermoelectric materials. Even if the Seebeck coefficient becomes large, a heat current inevitably accompanies a temperature gradient and thus makes a tradeoff. The main stream to prevail this issue is based on other materials with high power factor, such as doped narrow-gap semiconductors~\cite{Arita, Hamada, Zide}, or on nanostructuring, such as PbTe(1.5$~$nm)/Pb$_{0.927}$Eu$_{0.073}$Te(45~nm) multiple quantum well~\cite{Dresselhaus, Harman} and Bi$_{2}$Te$_{3}$~\cite{Venkatasubramanian}. It is well understood that this efficiency improvement is due to the sharp peaked electronic density of states (DOS) in low-dimensional materials~\cite{Dresselhaus, Bilc}, which is the optimal way toward high thermoelectric efficiency~\cite{Mahan}. Low dimensional systems could have dramatically larger $\mathcal{Z}T$ values than the corresponding bulk materials because of decreased thermal conductivity caused by phonon boundary scattering and improved power factors on account of quantum confinement. Although the efficiency is largely enhanced via dimensionality reduction, however, it typically affects electronic properties of conventional materials. Large efforts in improving thermoelectric performance target energy filtering, which provides a way to increase the Seebeck coefficient by introducing a strongly energy-dependent scattering mechanism~\cite{Mahan, Neophytou, Fomin, Zianni}. Recent advances in fabrication technologies have made exploring two-dimensional materials possible for thermoelectric applications~\cite{Wei, Zuev, Kato, Buscema, Konabe1}.
\par
Recently, isolated two-dimensional black phosphorus (BP), known as phosphorene with a puckered structure, received tremendous interest owing to its extraordinary electronic and optical properties in engineering applications~\cite{Liu, Das, Kamalakar, Nan}. The optical and transport properties of monolayer of BP exhibit strong in-plane anisotropy as bulk BP for two distinct zigzag and armchair directions. These anisotropic features mostly originate from anisotropic bands, like silicon~\cite{Markussen}. A nearly direct band gap of BP increases with decreasing number of layers from $0.3~e{\rm V}$ in bulk to $0.8~e{\rm V}<E_{g}<2~e{\rm V}$ for a monolayer~\cite{Takao, Keyes, Jiong, Katsnelson, Liangbo, Sahin, Tran, Ouyang}.
According to theoretical predictions, phosphorene has a high carrier mobility of around $1000~{\rm cm^2V^{-1}s^{-1}}$ ~\cite{Li} and a high on/off ratio of $10^4$ in phosphorene field-effect transistor at room temperature~\cite{Neal}. Besides the bulk BP, it has also predicted that the phosphorene may have unique potential thermoelectric applications~\cite{Zhang, Xu, Faghaninia, Cai, Flores, Jiang, Ong, Lv, Qin}.

\par
In this paper, we first propose an accurate low-energy model Hamiltonian protected all needed symmetries and compare that with those that appeared in literature. Then, we investigate the electronic contribution to the thermoelectric transport of the monolayer of phosphorene. We consider a phosphorene sheet in diffusive transport regime when thermal gradients and bias voltages are applied to the system. The generalized Boltzmann transport equation is applied to obtain the conductivity, Seebeck coefficient and the figure of merit. Moreover, the diffusive transport coefficients are calculated by considering a short-range potential and a long-range charge-charge Coulomb potential with a Thomas-Fermi screening as the source of scattering. Our calculations show that although the electrical conductivity of phosphorene is highly anisotropic, the Seebeck coefficient and the corresponding figure of merit are nearly isotropic. The figure of merit, which is a measure of thermoelectric efficiency, reaches to $\sim 1.2$ at low temperatures, irrespective of the underlying scattering mechanisms. We also investigate the effect of the interband coupling term on thermoelectric transport coefficients. These results propose that a monolayer of phosphorene could be a promising material for the thermoelectric applications.
\par
This paper is organized as follows. In Sec.~\ref{sec:model}, we first introduce the system and then explain the method which is used to calculate the conductivity and thermoelectric coefficients using the generalized Boltzmann method. In Sec.~\ref{sec:results}, we present and describe our numerical results for the conductivity and thermoelectric coefficients for phosphorene. Finally, we conclude and summarize our main results in Sec.~\ref{sec:concl}.

\begin{center}
\begin{table*}
\begin{tabular}{c|c|c|c|c|c|c|c|c|c|c|c} \hline
~~ & $E_g$ &~$m_{cx}$~ & ~$m_{vx}$~ & ~$m_{cy}$~ & ~$m_{vy}$~ & ~$\gamma$~ & ~$\beta$ &~$\eta_c$~&~ $\eta_v$~ &~ $\nu_c$~ &~ $\nu_v$ \\ \hline
Present & 0.912 & 0.146 & 0.131 & 1.240 & 7.857 &0.480 & 0 & 0.008 & 0.038 & 0.030 &0.005 \\ \hline
Ref. [\onlinecite{TL}] & 2.00 & 0.15 & 0.15 & 0.70 & 1.00 & 0.2839 & 0.0101& 0.2137& 0.2137& 0.0544 & 0.0381\\ \hline
Ref. [\onlinecite{castro}] & 0.70 & 0.1128 &  0.1080 & 1.5123 & $\infty$ & 0.4862 & 0.0353 & 0 & 0.0151&0.0252& 0 \\ \hline
\end{tabular}
\caption{The effective band masses, $m_{cx}$, $m_{cy}$, $m_{vx}$, and $m_{vy}$ (in units of the electron bare mass $m_0$), gap energy $E_{g}$ (in units of eV), and $\gamma=\vartheta a_x^0/ \pi$ (in units of eV nm) where $\vartheta=4$ or $6.85$, $\beta=\theta (a_x^0/\pi)^2$ (in units of eV nm$^2$) where $\theta=2$ or $7$, $\eta_s=\eta_0/m_{sx}-\gamma^2/E_g$ (in units of eV nm$^{2}$) and $\nu$ (in units of eV nm$^2$) based of theoretical works in Refs.~[\onlinecite{TL}] and [\onlinecite{castro}]. The values of $\vartheta$ and $\theta$ differ in those references. In this work we use parameters presented in [\onlinecite{Elahi}], the first row.}
\label{tab1}
\end{table*}
\end{center}

\section{Model and Basic Formalism}\label{sec:model}

\subsection{Hamiltonian of Monolayer Phosphorene}

Phosphorene has an orthorhombic puckered structure. The lattice constants of the conventional unit cell, considering four atoms per unit cell in $x$ (armchair) and $y$ (zigzag) directions, are respectively $a_x = 4.63$ \AA~and $a_y = 3.3$ \AA. Notice that the primitive unit cell's lattice constants are $a^0_{x/y} = a_{x/y}/2$. The spin degeneracy of the system is $g_s =2$ and possesses no valley degeneracy.

We consider a monolayer of phosphorene at low temperature. The electronic band structure of phosphorene has been calculated based on VASP package density-functional theory~\cite{Elahi}. The VASP package provides the first conduction band in the vicinity of the $\Gamma$ point, which is the exact position of the conduction band minimum. However, the VASP package suggests a slightly indirect band gap with its actual valance maximum occurring along the $\Gamma-Y$ high symmetry line~\cite{PLi}. Having ignored this slight shift, we can write down a low-energy model Hamiltonian. For this purpose, the electronic band structure basically could be described by a four-band model in the tight-binding model, however, it can be expressed by a two-band model owing to the C$_{2h}$ point group invariance. Expanding the tight-binding model~\cite{Ezawa, Katsnelson} around the $\Gamma$ point, one obtains the low-energy ${\bf k} \cdot {\bf p}$ model of phosphorene~\cite{castro} as,
\begin{equation}\label{Ham-rev}
H_{{\rm eff}}= \left(
\begin{array}{cc}
	E_c +  \eta_c k_x^2 + \nu_c k_y^2 & \quad
	\gamma k_x + \alpha k_{x}^{2} + \beta k_{y}^{2}\\
	\gamma k_x + \alpha k_{x}^{2} + \beta k_{y}^{2} & \quad
	E_v - \eta_v k_x^2 - \nu_v k_y^2
\end{array}
\right)
\end{equation}
in the conduction and valence band basis. Parameter $\alpha$ is usually ignored because of the existence of the linear leading order term $\gamma k_{x}$. Odd crosses terms of momentum components in the dispersion relation due to simultaneously nonzero $\gamma$ and $\beta$ break the time reversal (TR) symmetry. Within the L\"{o}wdin partitioning procedure the $\gamma k_{x}$ term comes from the unperturbed Hamiltonian~\cite{castro} and must valid. We obtain the TR invariant low-energy Hamiltonian of monolayer phosphorene as,
\begin{equation}\label{Ham}
H_{{\rm eff}}= \left(
\begin{array}{cc}
	E_c +  \eta_c k_x^2 + \nu_c k_y^2 & \quad
	\gamma k_x \\
	\gamma k_x & \quad
	E_v - \eta_v k_x^2 - \nu_v k_y^2
\end{array}
\right)
\end{equation}
where $E_{c(v)}$ is the band edge at the $\Gamma$ point with direct energy gap $E_g = E_c - E_v$ and the off-diagonal $\gamma k_x$ element is the interband coupling term with the real parameter $\gamma$. Other parameters can be extracted from the knowledge of DFT results~\cite{Elahi} where we have
$E_g=0.912$ eV, $\eta_c=0.008$, $\eta_v=0.038$ in units of eV nm$^{2}$, $\nu_c=0.030$ and $\nu_v=0.005$ in units of eV nm$^2$ which implies that effective masses have values $m_{cx}=0.146$, $m_{vx}=0.131$, $m_{cy}=1.240$ and $m_{vy}=7.857$  in units of the electron bare mass $m_0$. Notice that the hole mass along the zigzag ($y$ direction) is much (almost $10$ times) greater than that along the armchair ($x$ direction) which induces strong in-plane anisotropy. The only parameter which remains to identify is the $\gamma$, and we find $\gamma=0.480$ eV nm, by fitting the low-energy dispersion of the model Hamiltonian to that obtained by DFT-VASP results.
Notice that due to the time-reversal symmetry, the off-diagonal term includes only $\gamma k_x$ and other terms like $\beta k^2_y$ might be zero.

As discussed in Ref. [\onlinecite{Li}] based on the symmetry of the system, it is necessary to have a finite value of the $\nu_v$ albeit it is zero in [\onlinecite{castro}]. Remarkably, due to the time-reversal symmetry $\beta$ possibly being zero, however it is finite in the other parameterized Hamiltonian.

\begin{figure}
\centering
\includegraphics[width=3in]{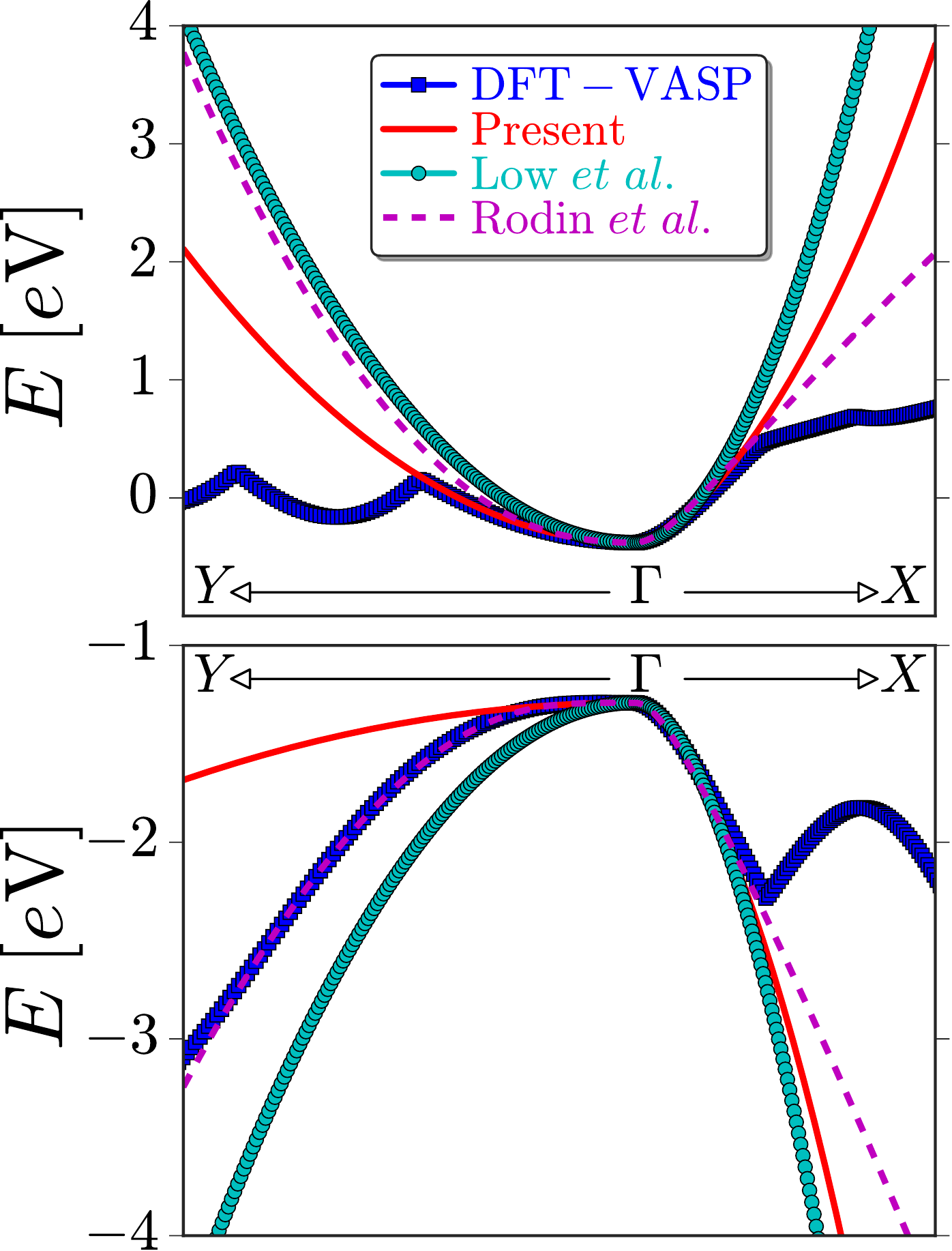}
\caption{(Color online) Phosphorene band energy dispersion along the $Y-\Gamma-X$ direction in the Brillouin zone. The conduction and valence bands are compared with those theoretical works in Refs.~[\onlinecite{TL}] and [\onlinecite{castro}] and with simulation results obtained within DFT-VASP (black curves) in Ref. [\onlinecite{Elahi}].}
\label{energyDispersion}
\end{figure}

\begin{figure}
\centering
\includegraphics[width=3in]{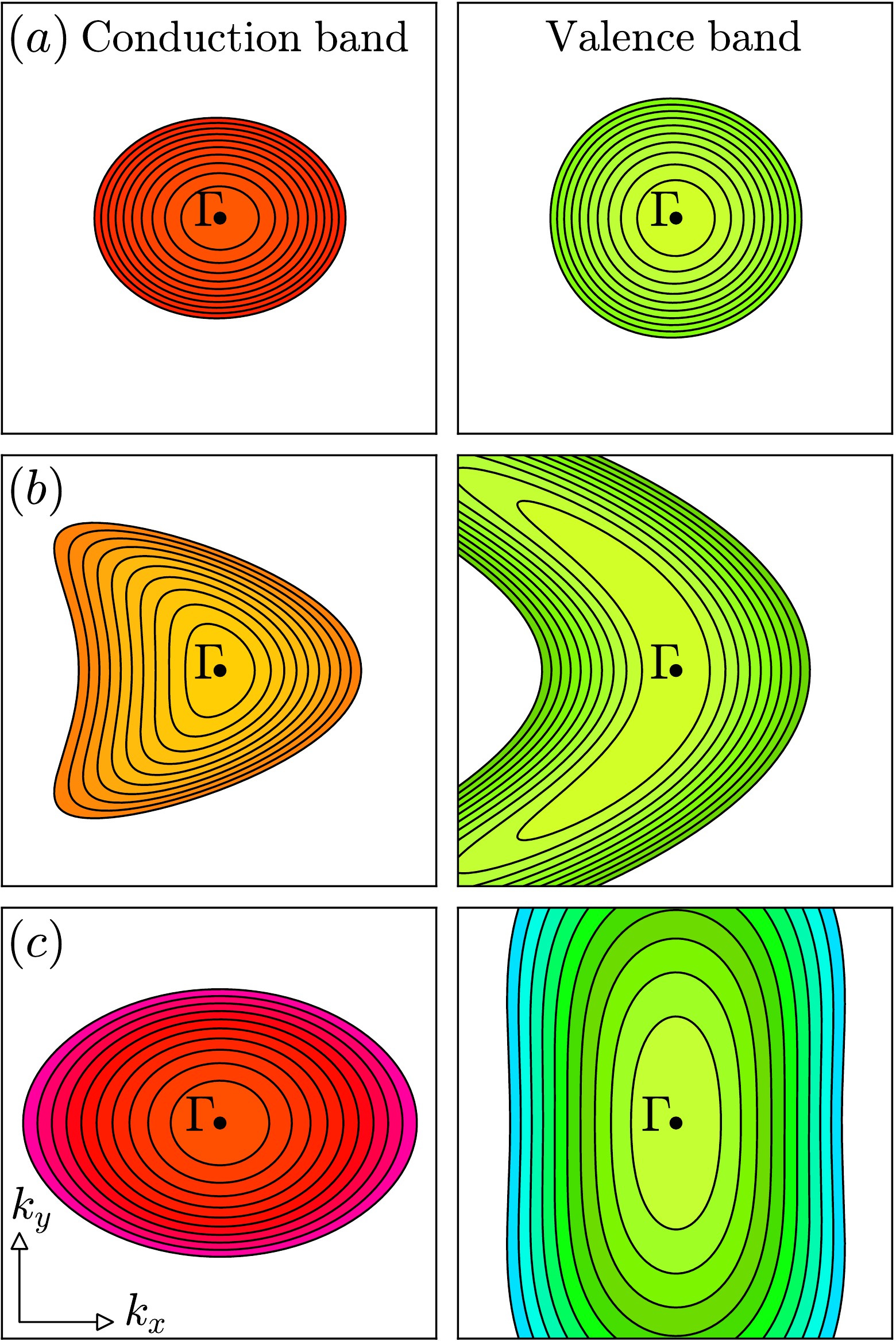}
\caption{(Color online) Isofrequency contour surface in the $k-$space at zero temperature for $E_{s}(\vec{k}_{\rm F})=0.05-0.5$ eV with a step of $0.05$ eV in both the electron and hole doped cases. The first and second rows (a) and (b) refer to parameters used in Refs.~[\onlinecite{TL}] and [\onlinecite{castro}], respectively. The last row, (c) plots are based on our model Hamiltonian.}
\label{contours}
\end{figure}

We demonstrate the band dispersion of the conduction (upper panel) and valence bands (lower panel) in Fig.~\ref{energyDispersion} where we compare our results with theoretical works in Refs.~[\onlinecite{castro}] and [\onlinecite{TL}]  and VASP-DFT simulations~\cite{Elahi}. All parameters are given in Table I. In the vicinity of the $\Gamma$ point, all discussed models capture the physics of the low energy along one direction of the momentum. In the 2D case, the isofrequency profiles are obtained by horizontally cutting the dispersion surface separately calculated by means of a plane wave model. We illustrate an isofrequency contour surface in the $k$ space to explore their symmetries for $E_{s}(\vec{k}_{\rm F})=0.05-0.5$ eV with a step of $0.05$ eV in both the electron and hole doped cases in Fig.~\ref{contours}. The first and second rows (a) and (b) refer to the Fermi surfaces with parameters used in Refs.~[\onlinecite{TL}] and [\onlinecite{castro}], respectively. As we stated before, the mass values in Ref.~[\onlinecite{TL}] are not entirely suitable for phosphorene, although the isofrequency counter Fermi surfaces are quite like elliptic structure for $E_{s}(\vec{k}_{\rm F})<0.5$ eV and predict that the interband coupling term can be ignored. This is also the case in the dispersion relation structure of proposed low energy Hamiltonian in [\onlinecite{Katsnelson}]. On the other hand, the counter plot of  Ref.~[\onlinecite{castro}] breaks the time-reveal symmetry even at low electron or hole density. This is basically based on the extra off-diagonal terms that appeared in the low-energy of their model Hamiltonian. Finally, we present the isofrequency counter surface in the $k$ space based on our parameters and importantly the shape of the Fermi surfaces in our model are almost an elliptic shape especially at low charge density. Our results predict that the interband coupling term plays a role.

By diagonalizing the Hamiltonian Eq. (\ref{Ham}), we end up with two energy bands given by,
\begin{eqnarray}
E_{\tau}=\frac{1}{2}[H_c + H_v + \tau\sqrt{4H_{cv}^{2}+(H_{c}-H_{v})^{2}}]\label{disp}
\end{eqnarray}
with $H_{c}=E_{c}+\eta_{c}k_{x}^{2}+\nu_{c}k_{y}^{2}$, $H_{v}=E_{v}-\eta_{v}k_{x}^{2}-\nu_{v}k_{y}^{2}$, $H_{cv}=\gamma k_{x}$, and $\tau=\pm 1$ denotes the conduction (valence) band. The corresponding eigenvector reads
\begin{eqnarray}
\Psi_{c(v)}=\frac{1}{\sqrt{1+|\chi_{c(v)}|^2}}
\begin{pmatrix}
\chi_{c(v)}\\ 1
\end{pmatrix}
\end{eqnarray}
where $\chi_{c(v)}=[H_{c}-H_{v}+\tau\sqrt{4H_{cv}^{2}+(H_{c}-H_{v})^{2}}]/2 H_{cv}$.
\par
Furthermore, having calculated the band energy dispersion given by Eq. (\ref{disp}), the $x$ and $y$ components of the velocity can be calculated as
\begin{eqnarray}
v_{x}&=&k_{x}\left[\eta_{c}-\eta_{v}+\tau\frac{2\gamma^{2}+(H_{c}-H_{v})(\eta_{c}+\eta_{v})}{\sqrt{4H_{cv}^{2}+(H_{c}-H_{v})^{2}}}\right]\\
v_{y}&=&k_{y}\left[\nu_{c}-\nu_{v}+\tau\frac{(H_{c}-H_{v})(\nu_{c}+\nu_{v})}{\sqrt{4H_{cv}^{2}+(H_{c}-H_{v})^{2}}}\right]
\end{eqnarray}

\subsection{Anisotropic transport framework}

In this section we use the generalized semiclassical Boltzmann
formalism for an anisotropic system to establish the transport coefficients in the diffusive regime. In particular, we take into account two important cases of short-range (SR) impurities (\textit{e.g.}, defects or neutral adatoms) with Dirac delta potential and long-range (LR) Coulomb impurities
in our investigation. The thermoelectric properties of phosphorene in the presence of both the electric field and the temperature gradient will be found.

In the diffusive regime, the transport coefficients can
be obtained from the charge current and the energy flux density. More details are provided in Appendix A. The nonequilibrium distribution function in the presence of driving forces is needed to calculate the current densities. For this purpose, we take the Boltzmann equation up to a linear order in the presence of thermoelectric fields. The collision integral is given by
\begin{eqnarray}
\left(\frac{df}{dt}\right)_{{\rm coll.}}=\int\frac{d^2k'}{(2\pi)^2}w({\bf k},{\bf k^{\prime}})\left[f({\bf k}, \bm{\mathcal{E}}, T)-f({\bf k^{\prime}}, \bm{\mathcal{E}}, T)\right]\nonumber\\
\label{eq4}
\end{eqnarray}
where $w({\bf k},{\bf k^{\prime}})$ is the scattering rate from state ${\bf k}$ to state ${\bf k^{\prime}}$ which needs to be specified according to the microscopic origin of the scattering mechanisms. As the relaxation time approximation provides an inadequate explanation for the full aspects of the anisotropic features of the transport properties, an exact integral equation approach might be implemented~\cite{Vyborny, zare2016, faridi}. The scattering $w({\bf k}, {\bf k^{\prime}})$ rates using the Fermi golden rule within the lowest order of the Born approximation are given by
\begin{eqnarray}
w({\bf k}, {\bf k'})=\frac{2\pi}{\hbar}n_{{\rm imp}}\big\vert\langle {\bf k'}|\hat{V}|{\bf k}\rangle\big\vert^2 \delta(\varepsilon_{\bf k}-\varepsilon_{\bf k'})
\end{eqnarray}
where $n_{{\rm imp}}$ is the areal density of randomly distributed scatterers and $\hat{V}_{{\bf k}-{\bf k^{\prime}}}$ is the Fourier transformation of the interaction potential between an electron and a single impurity. The short-ranged impurities are approximated with a zero-range hard-core potential $\hat{V}_{{\bf k}-{\bf k^{\prime}}}=V_{0}$. On the other hand, the long-ranged Coulombic interaction owing to the charged impurities is screened by other electrons of
the system like the Thomas-Fermi approach.
The generalized conductivity $\sigma(\varepsilon; \theta, \theta')$ is given by
\begin{eqnarray}
\sigma(\varepsilon; \theta, \theta')&=&e^{2}\int\frac{d^{2}k}{(2\pi)^{2}}\delta\left(\varepsilon-\varepsilon({\bf k})\right) v^2(\phi)\nonumber\\
&&\quad\left[a(\phi)\cos\theta+b(\phi)\sin\theta\right]\cos(\theta-\xi(\phi))\qquad
\end{eqnarray}
with $\theta=\theta^{\prime}=0$ for $\sigma_{xx}$ and $\theta=\theta^{\prime}=\pi/2$ for $\sigma_{yy}$. We concentrate on low enough temperatures where only electrons contribute effectively in thermal transport and disregard phonon contribution.

\begin{figure}
\includegraphics[width=8.5cm]{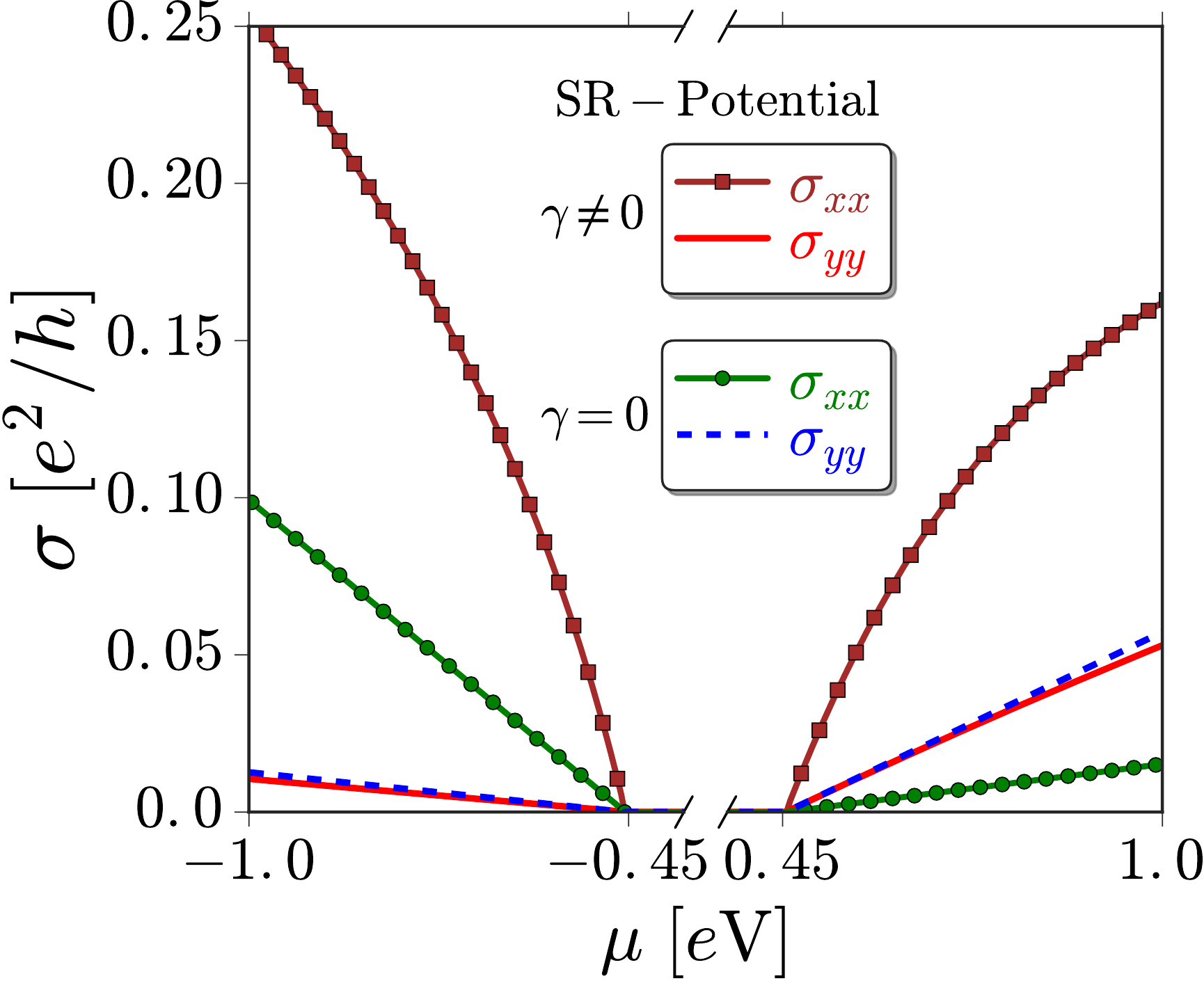}
\caption{(Color online) The conductivity of monolayer phosphorene as a function of the chemical potential $\mu$ at the presence of short-range impurity potential along the zigzag, $\sigma_{yy}$, and armchair, $\sigma_{xx}$, directions. The effect of the coupling term $\gamma$ is also shown. Note that the chemical potential is measured from the middle of the gap value.}
\label{fig3}
\end{figure}

\begin{figure}
\includegraphics[width=8.4cm]{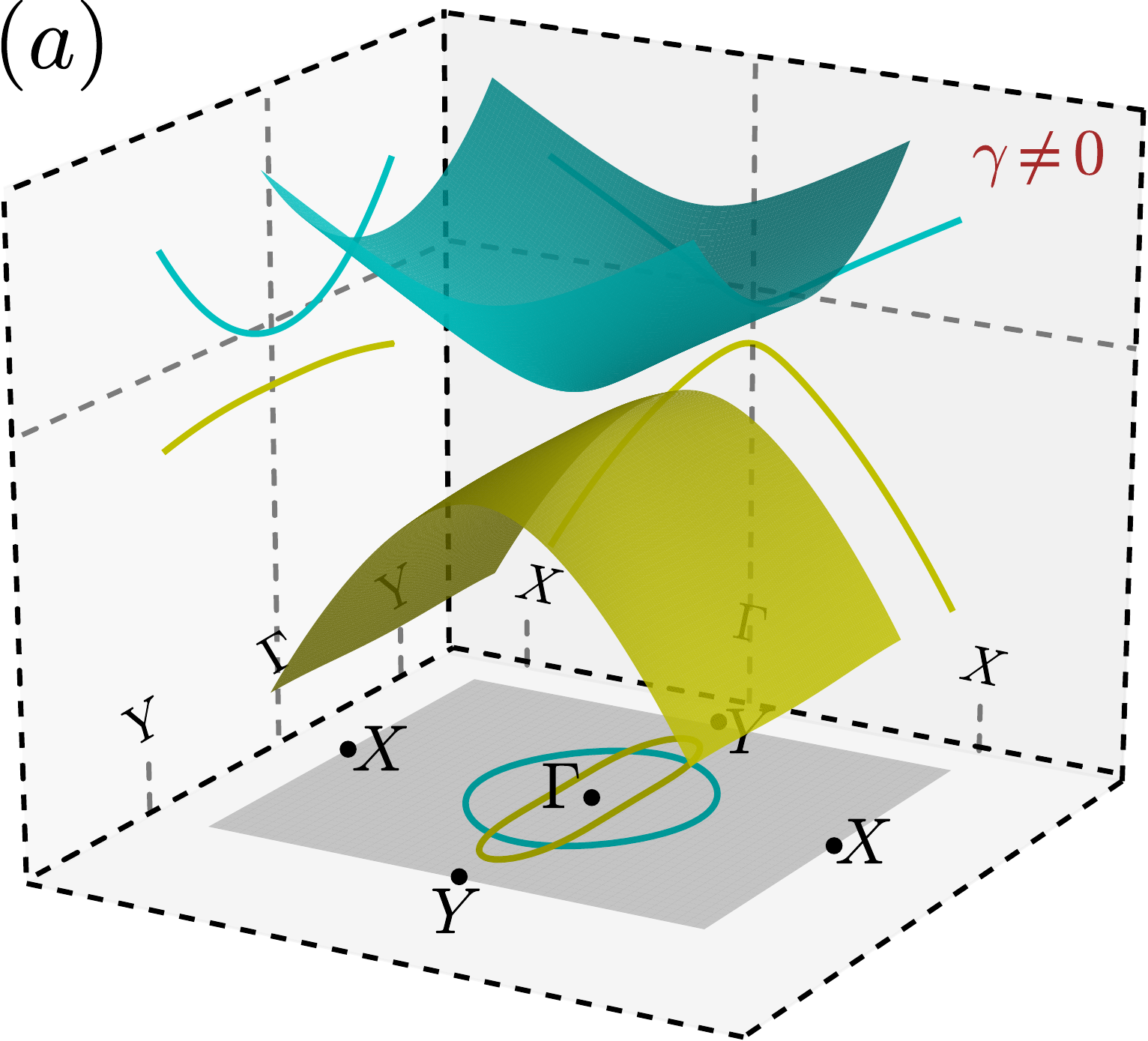}\\
\includegraphics[width=8.4cm]{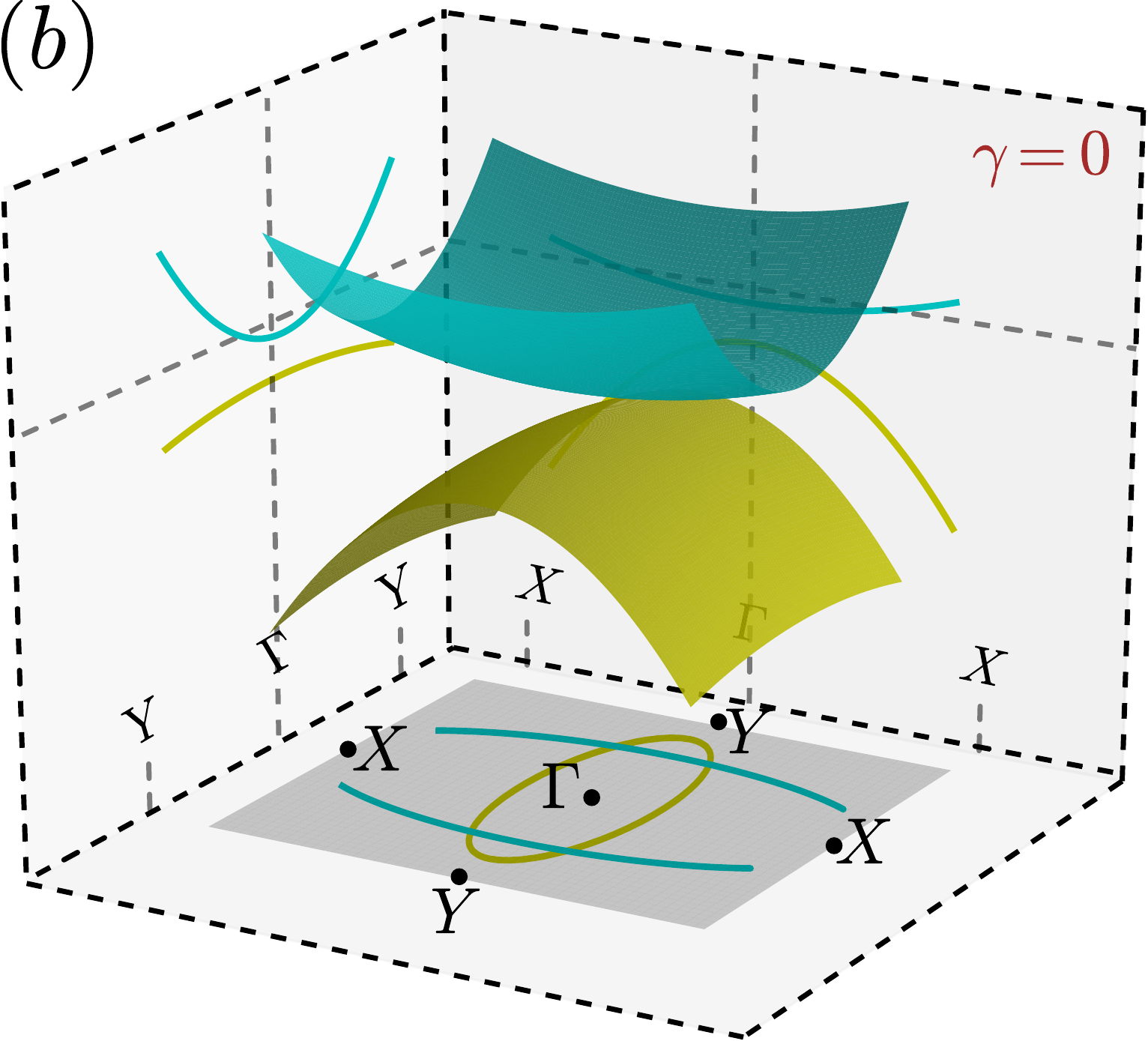}
\caption{(Color online) The band structure of phosphorene in the first Brillouin zone, indicated by the gray square plane. Dispersion of phosphorene is depicted in the planes of $X-\Gamma-X$ and $Y-\Gamma-Y$. The corresponding constant-energy contour plots in the Brillouin zone are shown in both $n$- and $p$-doped cases. The influence of the coupling term $\gamma$ is also indicated in panels (a) and (b).}\label{fig4}
\end{figure}

\section{Numerical Results and Discussion}\label{sec:results}

In this section our numerical results for the thermoelectric transport in phosphorene are presented. We investigate the electrical conductivity, Seebeck coefficient ($\mathcal{S}$), and its corresponding figure of merit $\mathcal{Z}T$, considering both the SR and LR potentials. It should be noted that, we set $T\sim20~{\rm K}$ in all calculated quantities. Moreover, $n_{{\rm imp.}}=10^{10}{\rm cm^{-2}}$ corresponding to the chemical potential approximately $\mu\sim10^{-4}~e{\rm V}$, is used for the impurity concentration of both short-range and long-range potentials to ensure that the diluteness criteria is satisfied. It is worthwhile to mention that there are essential criteria for utilizing the Boltzmann equation. These criteria are listed as follows. Particles might interact via binary collisions, impurity density is low in terms of the charge carriers, an external field might has long-range wavelength and all collisions are elastic and involve only uncorrelated particles.
\par
Figure (\ref{fig3}) shows the variation of the electrical conductivity of phosphorene versus the chemical potential $\mu$ in the presence of short-range impurity interaction with $V_{{\bf k}- {\bf k^{\prime}}}=V_{0}=1000~e{\rm V}$\AA$^{2}$~\cite{sarma}, in the zigzag, $\sigma_{yy}$, and armchair, $\sigma_{xx}$, directions. The
influence of the interband coupling term $\gamma$ is also indicated. While at the presence of the coupling term $\gamma$ the electrical conductivity in the armchair direction $\sigma_{xx}$ is greater than the conductivity in the zigzag direction $\sigma_{yy}$, for both $n$- and $p$-doped regimes, however the conductivity in the armchair direction is smaller than that of the zigzag direction $\sigma_{xx}<\sigma_{yy}$ for the $n$-doped regime. Intriguingly, the conductivity in the armchair direction $\sigma_{xx}$ is more influenced by the inclusion of the coupling term $\gamma$ and enhanced significantly. In both doping regimes, the coupling term does not alter notably the conductivity in the zigzag direction $\sigma_{yy}$. All these behaviors are the characteristics of the dispersion of monolayer phosphorene, as indicated in Fig. (\ref{fig4}). In the absence of the coupling term $\gamma=0$, phosphorene dispersion relation reduces to two separate ovals for the conduction and valence bands. In order to understand aforementioned features, we use the intuition based on the Drude formula with an effective mass tensor, $\sigma\sim1/{\bf m}^{*}$, of the transport. Around the $\Gamma$ point, the components of the effective mass for $\gamma\neq0$ are $m_{c,xx}^{-1} = 0.333>m_{c,yy}^{-1}= 0.06145$, and $m_{v,yy}^{-1} = -0.00969<<m_{v,xx}^{-1} = -0.393$, while for the $\gamma=0$ are $m_{c,xx}^{-1}= 0.01637<m_{c,yy}^{-1}= 0.061452$, and $m_{v,xx}^{-1} = -0.07613>m_{v,yy}^{-1} = -0.00969$. It is worth noting that unlike the graphene~\cite{zare2015}, the conductivity of phosphorene has an explicit energy dependence when only short-range scatterers are present.
\begin{figure}
\includegraphics[width=8.5cm]{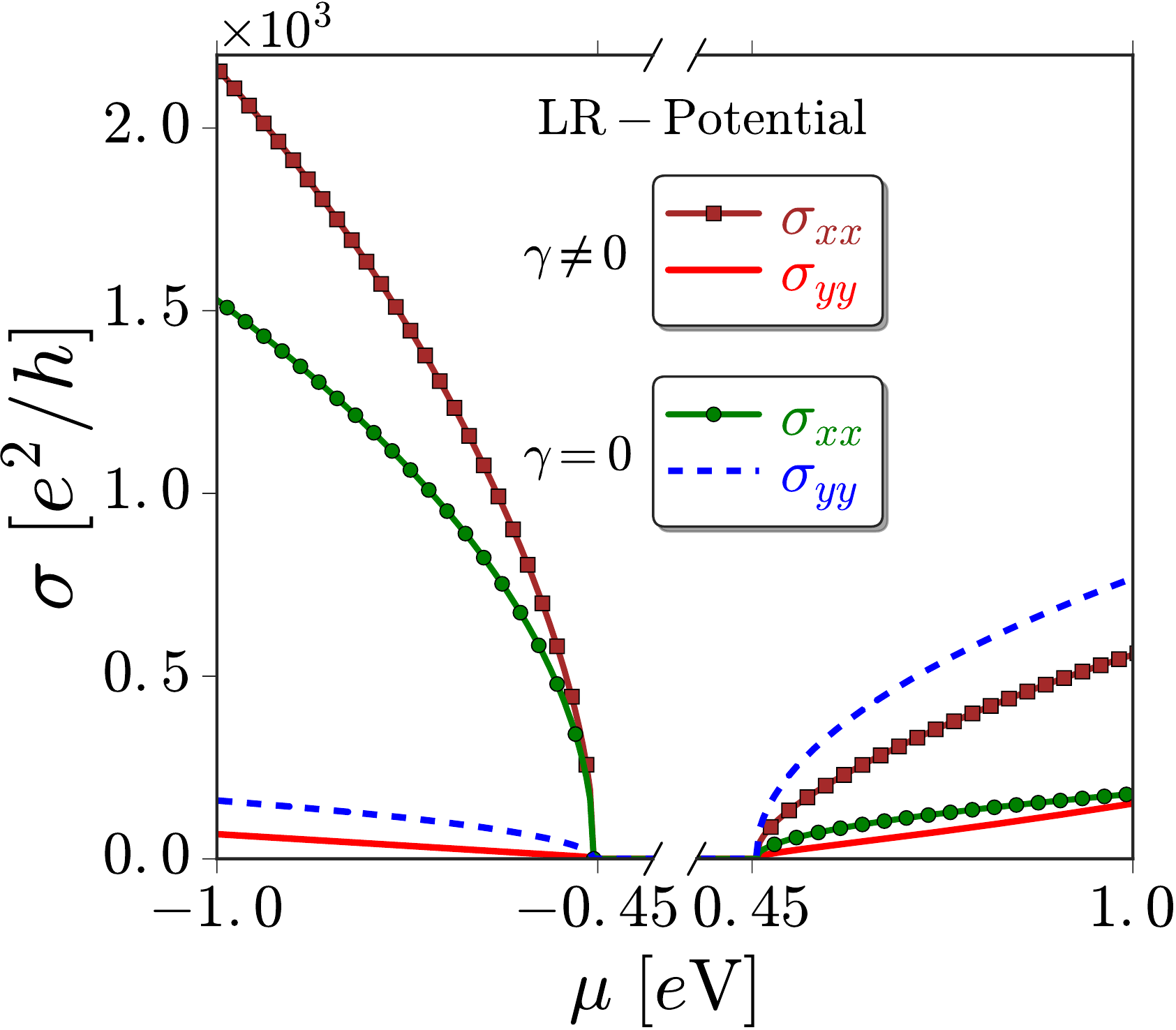}
\caption{(Color online) The conductivity of monolayer phosphorene as a function of doping $\mu$ at the presence of long-range charged impurity potential for the zigzag, $\sigma_{yy}$, and armchair, $\sigma_{xx}$, directions. The effect of coupling term $\gamma$ is also shown.}
\label{fig5}
\end{figure}

\begin{figure}
\includegraphics[width=8.5cm]{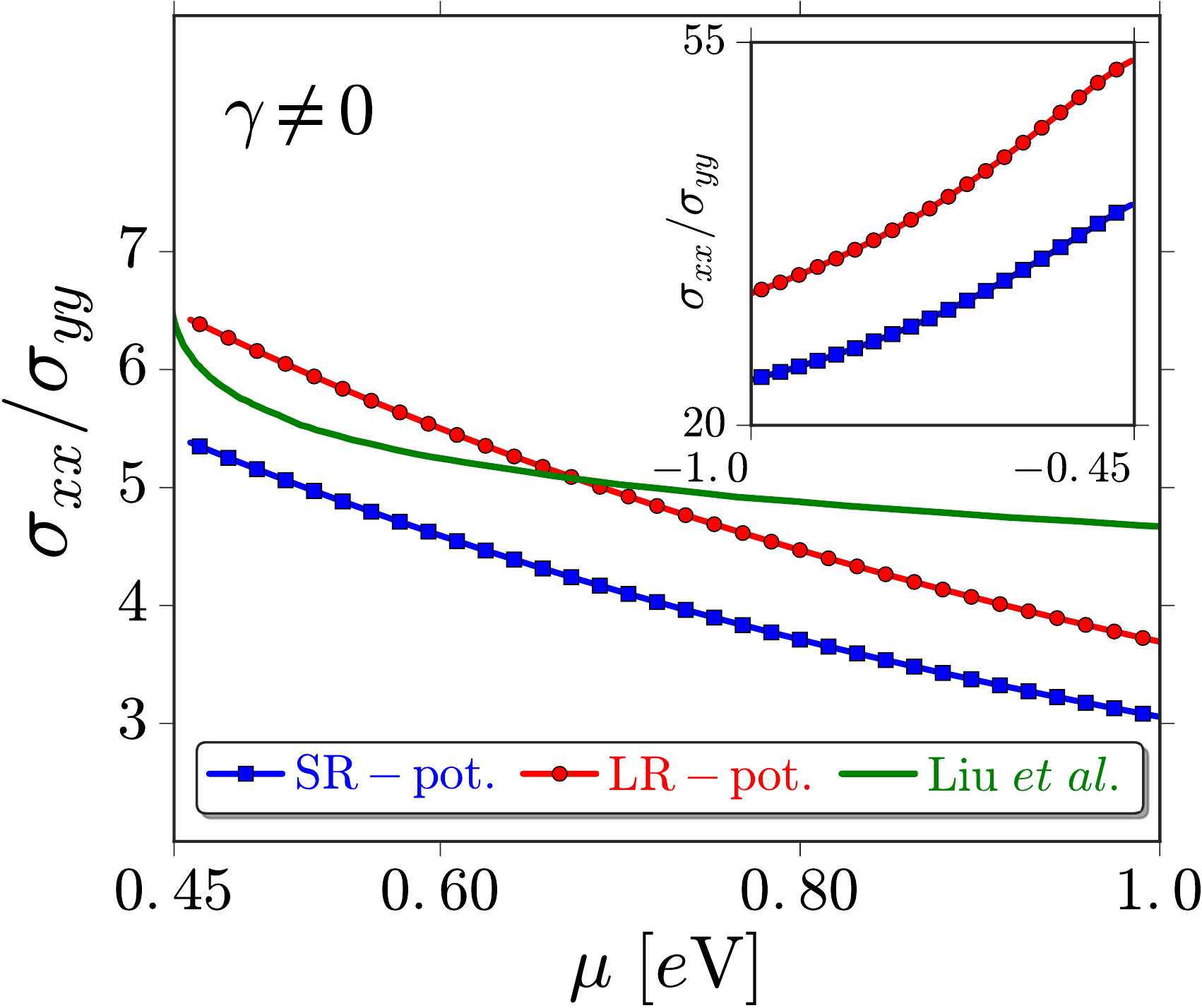}
\caption{(Color online) The anisotropy ratio of the conductivity, $\sigma_{xx}/\sigma_{yy}$, of monolayer phosphorene as a function of electron doping $\mu$ at the presence of short- and long-range impurity potentials. Inset: the ratio of the conductivity as a function of the hole chemical potential. Solid line refers to data calculated in [Ref.\onlinecite{Li1}] for long-ranged impurity potential at $d=0$.}
\label{fig6}
\end{figure}
\par
Due to the fact that the long-range charge-impurity
Coulomb interactions are mostly the dominant scatterers in samples, we also consider the Coulomb interaction. To this end, we use an interaction potential including static Thomas-Fermi screening, as is commonly used for a 2D electron gas~\cite{ando} to account partially for screening, as $V_{{\bf k}- {\bf k^{\prime}}}=2\pi e^{2}/(\varepsilon\left(\vert{\bf k}-{\bf k^{\prime}}\vert+q_{_{{\rm TF}}}\right))$
where $q_{_{{\rm TF}}}=2\pi e^{2}N(\mu)/\varepsilon$ is the Thomas-Fermi screening vector with the density of states of the system, $N(\mu)$. We use the dielectric constant of the common substrate SiO$_{2}$ which is about $\varepsilon=2.45$. In Fig. (\ref{fig5}), the conductivity of phosphorene as a function of doping is plotted in the presence of LR potentials, for both the zigzag, $\sigma_{yy}$, and armchair, $\sigma_{xx}$, directions. The overall energy dependence is the same as SR interactions which indicated that despite the details of scattering phenomena, the dispersion of phosphorene plays a main role in the conductivity. However, there is a clear discrepancy between two scattering mechanisms for the role of the coupling parameter $\gamma$. Interestingly, the conductivity in the zigzag direction $\sigma_{yy}$ is more affected by the coupling term than the $\sigma_{xx}$, in contrary to SR interactions where $\sigma_{xx}$ is more altered by the coupling term. On the other hand, while the $\sigma_{xx}$ is enhanced by the coupling for SR potentials, here $\sigma_{yy}$ is suppressed due to the coupling term. We should mention that at very low temperatures the variation of thermal conductivity will be similar to the charge conductivity $\mathcal{K}\approx (\pi^{2}/3)k_{{\rm B}}T \sigma$.
\begin{figure}
\includegraphics[width=8.5cm]{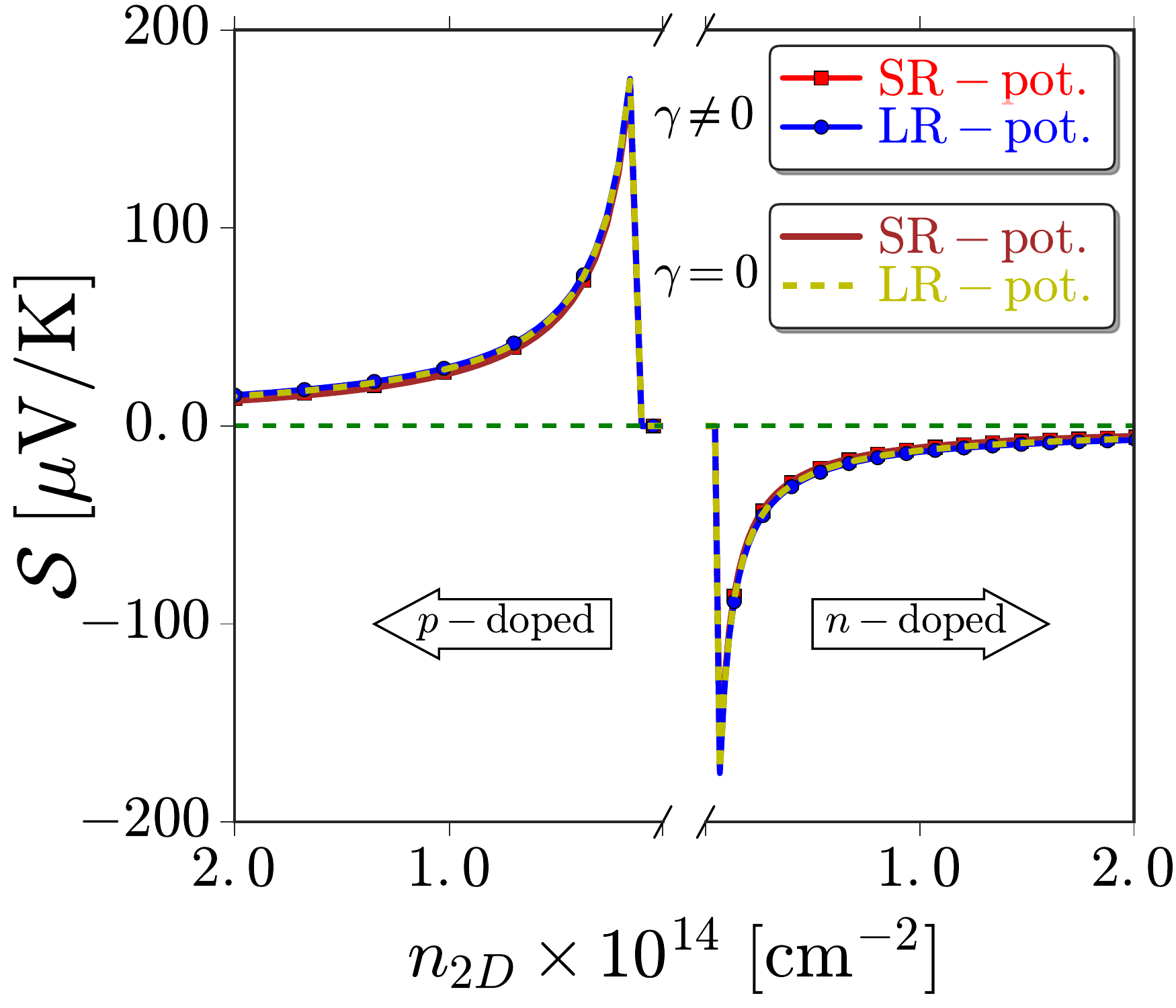}
\caption{(Color online) Seebeck coefficient of phosphorene as a function of carrier density $n_{2D}$ at the presence of short-
and long-range Coulomb potentials. The effect of the coupling term $\gamma$ is also depicted. Despite the $\gamma$, the Seebeck coefficients are nearly isotropic for both directions.}
\label{fig7}
\end{figure}

Figure~(\ref{fig6}) shows the anisotropy ratio calculated using the SR and LR potentials. First of all, as seen in the figure, the ratio is significantly large specially at low charge carrier density. The curves are monotonic in terms of the chemical potential and find that the ratio slightly changes with the type of impurity. In the inset, we show the anisotropic ratio of the hole doped case. We also compare our numerical results with those obtained by Liu \textit{et al},~\cite{Li1} in the case that $d=0$, the distance between charged impurity with phosphorene, in which they computed the mobility within the Boltzmann transport equation under detailed balance condition together with the anisotropy in momentum. As seen in the figure, there is a discrepancy between our fully self-consistent method with the approximated relaxation time result. This predicts that the Boltzmann transport equation with the anisotropic momentum can not provide a full description of the transport properties in phosphorene, except at very low doping regime.
\begin{figure}
\includegraphics[width=8.5cm]{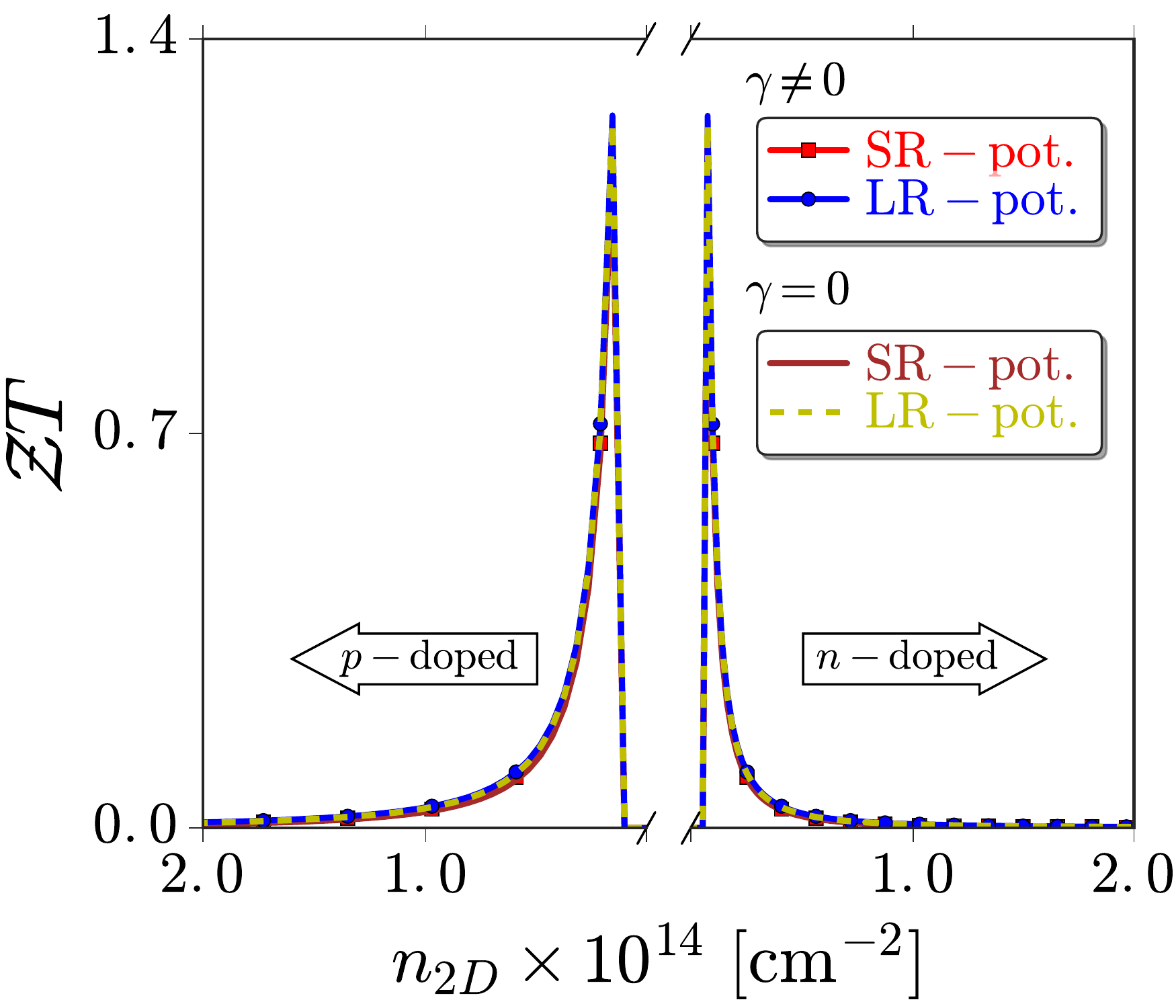}
\caption{(Color online) The variation of corresponding figures
of merit are depicted as a function of the carrier density at the presence of short- and long-range charge-impurity
potentials. The effect of the coupling term $\gamma$ is also depicted. Figures of merit are also nearly isotropic.}
\label{fig8}
\end{figure}
\par
\begin{figure}
\includegraphics[width=8.5cm]{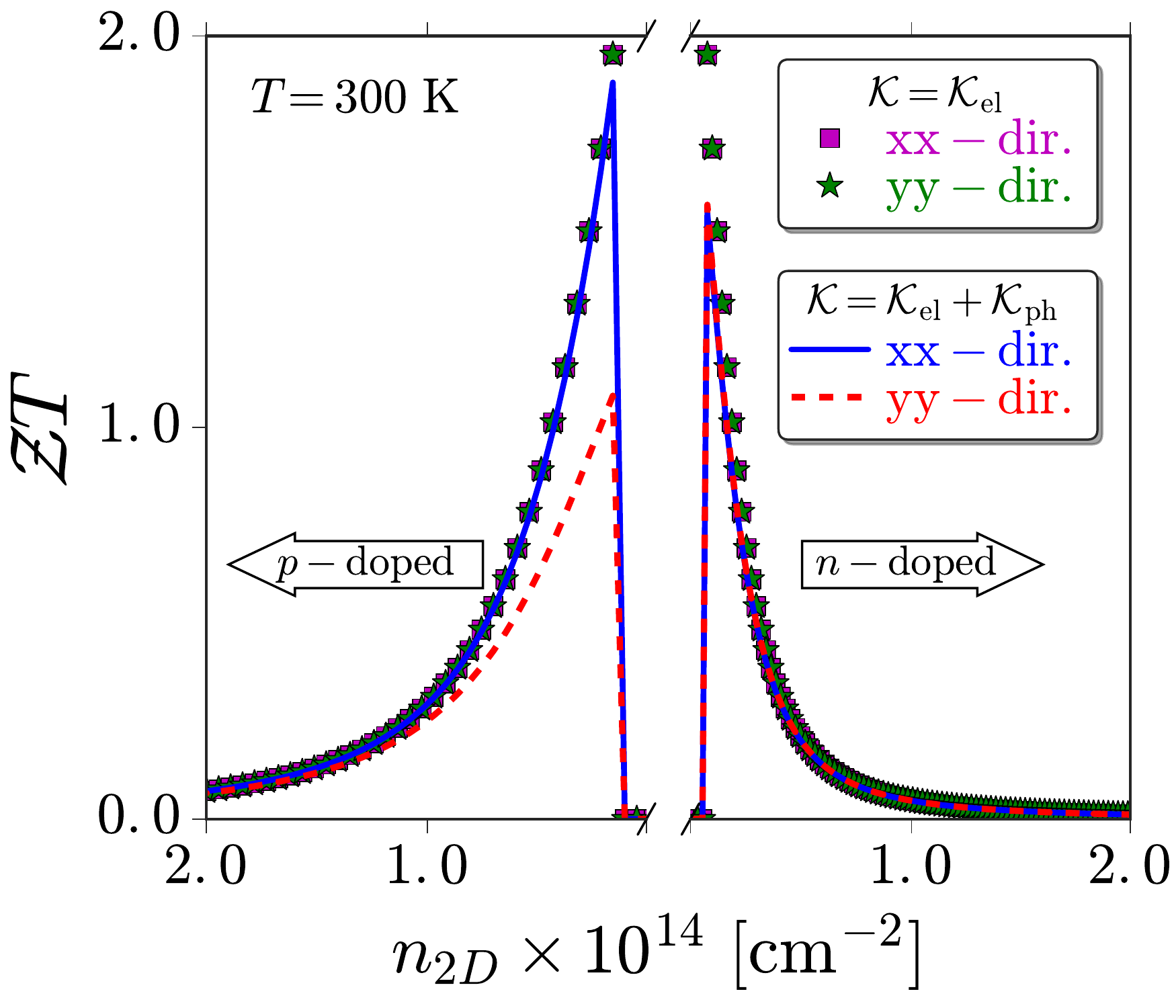}
\caption{(Color online) The variation of figures
of merit along the armchair and zigzag directions as a function of the carrier density at the presence of long-range charge-impurity potential at room temperature $T=300$~K. The effect of only electron contribution in the thermal conductivity $\mathcal{K}_{{\rm el}}$ is illustrated by symbols while the full effect of the electron and phonon contributions in the thermal conductivity $\mathcal{K}_{{\rm el}}+\mathcal{K}_{{\rm ph}}$ (which is $\sim13$ and $\sim30$ Wm$^{-1}$K$^{-1}$ along armchair and zigzag, respectively) are shown by solid and dashed lines.}
\label{fig9}
\end{figure}

The variation of  the Seebeck coefficients (thermopower) $\mathcal{S}$, as a more feasible quantity in real experiments, with doping at the presence of SR and LR potentials is obtained as shown in Fig. (\ref{fig7}). When the Fermi level lies in the valence band thermally activated holes, which move along the same direction as the temperature gradient owing to the positive charge, it results in a positive thermopower, however thermally activated electrons in the conduction band lead to a negative thermopower. Moreover, the figure of merit attains its maximum value around the chemical potential as can be seen in Fig. (\ref{fig8}) where the $\mathcal{Z}T$ is depicted as a function of the carrier density $n_{2D}$ for both scatterers. The figure of merit becomes large where the power factor $\mathcal{S}^{2}\sigma$ is very strong while the thermal transport $\mathcal{K}$ is not. Our results reveal that, in contrast to highly anisotropic electrical and thermal conductivities, the Seebeck coefficient and the thermoelectric figures of merit are nearly isotropic, consistent with the prior work~\cite{Faghaninia}. In fact, for the charge carrier density less than about $1\times10^{14}$~cm$^{-2}$, both $\sigma$ and its derivative with respect to the energy have approximately the same anisotropic behavior, consequently it leads to a nearly isotropic behavior of the Seebeck coefficient. Interestingly, in spite of underlying scattering mechanisms, thermopower and its corresponding figure of merit are not affected by the coupling term $\gamma$, on the contrary the charge conductivity. In Fig. (\ref{fig9}), the figures of merit, as a function of doping are plotted at $T=300$~K in the presence of LR potentials, for both the zigzag and armchair directions. Notice that in this case, we calculate Eq. (\ref{cond_ene}) in the Appendix numerically. The contribution of the phonon in the thermal conductivity $\mathcal{K}_{{\rm ph}}\sim$ 20-40 Wm$^{-1}$K$^{-1}$~\cite{Faghaninia, GQin, THLiu, YHong} can only affect the figure of merit and the thermopower is not altered by the presence of phonon. At high temperatures, the phonon becomes important but it only results in the overall decline of the figures of merit, without affecting their qualitative behavior.

\section{Conclusion}\label{sec:concl}

In conclusion, the thermoelectric transport in phosphorene in the presence of short- and long-ranged charged impurity potentials is studied using the generalized semiclassical Boltzmann approach for anisotropic systems. The charge conductivity, which is slightly different for $n$- and $p$-doped cases mostly owing to the unique dispersion of phosphorene, is found to be highly anisotropic, while the Seebeck coefficient and the corresponding figure of merit, without being affected either by type of scatterers or the presence/absence of coupling term, are nearly isotropic. Intriguingly, the conductivity for $n$-doped cases is more influenced by the coupling term, albeit in a dissimilar manner for different scatterers, than $p$-doped cases. Furthermore, it is shown that thermopower changes sign due to the conversion of electrons to holes and vice versa at the edge of the bands. We also reveal that phosphorene could be a very promising material for thermoelectric studies and applications.

\begin{table}
\begin{center}
\begin{tabular}{|l|l|l|l|l|l}
\hline
 & ~~~$T$ [K] & ~~$\mathcal{S}[\mu$VK$^{-1}$] & ~~~~$\mathcal{Z}T$ \\\hline
Phosphorene~\cite{Lv} & 300 & 3000 & 1.78 \\\hline
Phosphorene~\cite{Faghaninia} & 300, 500 & 2000, 2800 & 1.5, 3.8 \\\hline
Phosphorene~\cite{Shao2} & 300 & 500, 600 & 1.65, 2.12 \\\hline
Phosphorene~\cite{Dresselhaus2} & 300, 500 & 450, 500 & 0.1, 0.14 \\\hline
Phosphorene~\cite{Zhang} & 300 & 1400 & up to 6.5 \\\hline
Silicene~\cite{Pan} & 300 - 600 & up to 858 & 2.8 - 4.9 \\\hline
Graphene & 300 & up to 80 & 0.79 - 1 \\
\cite{Zuev, Wei, Checkelsky, Mazzamuto} &  &  &  \\\hline
Graphene~\cite{Zhong} & 150, 300 & up to 60, 120 & -- \\\hline
Graphene~\cite{Bao} & T<40, 300 & up to 12, 50 & -- \\\hline
MoS$_{2}$~\cite{Buscema} & 300, 500 & up to 10$^{5}$ & 0.5, 1 \\\hline
Present & 20 & $\sim$175 & $\sim$1.2 \\\hline
\end{tabular}
\caption{Reported values of the Seebeck coefficients and their corresponding figures of merit for monolayer of phosphorene, silicene, graphene, and MoS$_2$ systems.}\label{Tab:2}
\end{center}
\end{table}
\par
Since several works on thermodynamics in 2D crystalline material systems are available, a proper comparison with those results seems to be in order. Recent investigations, based on DFT calculations, showed that the $\mathcal{Z}T$ value of BP can only reach 0.22 at room temperature, while for monolayer phosphorene it reaches to 1.78~\cite{Lv}. It has been argued that applying strain is a practical way to enhance the thermoelectric efficiency of BP, and the largest $\mathcal{Z}T$ value of 0.87 can be achieved~\cite{Qin}. Although the $\mathcal{Z}T$ values obtained for BP are very small to compete with typical thermoelectric materials, e.g., Bi$_{2}$Te$_{3}$, Fei \textit{et al.} using first principle simulations showed that both the electrical and thermal conductance of monolayer phosphorene are highly anisotropic and the $\mathcal{Z}T$ value is greater than $1.0$ at room temperature and can attain up to $2.5$ at $500~{\rm K}$~\cite{Faghaninia}, due to the optimal ratio of conductances with orthogonally preferred conducting directions. Zhang \textit{et al.} demonstrated, based on first principle calculations, that the $\mathcal{Z}T$ value for phosphorene nanoribbons can achieve up to $6.4$ at room temperature~\cite{Zhang}. Liao \textit{et al.} implying first principle calculations reported $\mathcal{Z}T$ values of $0.1$, $0.14$ at $300$, $500~{\rm K}$, respectively, for $p$-doped samples~\cite{Dresselhaus2}. Lv \textit{et al.}, based on the semiclassical Boltzmann equation and DFT calculations, showed that $\mathcal{Z}T$ value of phosphorene at room temperature by strain can reach up to $1.65$~\cite{Shao2}.
\par
Pan \textit{et al.}, using the nonequilibrium Green’s function method and molecular dynamics simulations, predicted that the $\mathcal{Z}T$ value of zigzag silicene nanoribbon can achieve up to $4.9$~\cite{Pan}. Experimentally, the thermopower of graphene has been varied from 20 to 90 $\mu$VK$^{-1}$ while the temperature is changed from 10 to 300 K~\cite{Zuev}. Wei \textit{et al.}, experimentally achieved up to 50 $\mu$VK$^{-1}$ for the thermopower of graphene at a temperature range of 11–255 K~\cite{Wei}. Checkelsky \textit{et al.} reported measurement of thermopower in graphene that reaches up to 100 $\mu$VK$^{-1}$ at room temperature~\cite{Checkelsky}. By means of atomistic simulation, Mazzamuto \textit{et al.} predicted that the thermopower of graphene nanoribbons attain the value of 300 $\mu$VK$^{-1}$~\cite{Mazzamuto}. The investigation based on self-consistent Born approximation predicted the value of 0.4 $\mu$VK$^{-2}$ for the thermoelectric power of graphene $\mathcal{S}/T$, at the presence of charged impurity scatterers~\cite{Zhong}. Bao \textit{et al.,} by presenting a balance-equation-based theoretical examination of thermoelectric power in graphene, found that $\mathcal{S}$ changes from 1 to 50$\mu$VK$^{-1}$ as temperature goes from 10 to 300 K~\cite{Bao}. Buscema \textit{et al.}, by scanning photocurrent microscopy, have observed a thermopower as high as 10$^5$ $\mu$VK$^{-1}$ for a single-layer MoS$_{2}$, which is tunable via an external electric field~\cite{Buscema}. Finally, the predicted and measured values of the Seebeck coefficient and $\mathcal{Z}T$ of 2D materials are listed in Table~\ref{Tab:2}.

\par
\section{acknowledgments}
This work was partially supported by Iran Science Elites Federation.

\section*{appendix}

In the diffusive regime, the transport coefficients can
be obtained from the following expression for the charge current ${\bf j}$ and energy flux density ${\bf j}^{q}$,
\begin{eqnarray}
\left[\begin{array}{c}
{\bf j}  \\
{\bf j}^{q}
\end{array}
\right]=\int\frac{d^2k}{(2\pi)^2}
\left[\begin{array}{c}
-e  \\ \varepsilon({\bf k})-\mu
\end{array}
\right]{\bf v(k)} f({\bf k})
\label{currents}
\end{eqnarray}
where ${\bf v(k)}$ is the semiclassical velocity of the carriers which is related to the energy dispersion
$\varepsilon_{{\bf k}}$ through ${\bf v}=(1/\hbar)\nabla_{{\bf k}}\varepsilon_{{\bf k}}$. The nonequilibrium distribution function $f({\bf k})$ describes the evolution of the charge distribution in the presence of thermoelectric forces. In the linear response theory we seek a solution of Eq. (6) in the form of
\begin{eqnarray}
f({\bf k}, \bm{\mathcal{E}}, T)-f_{0}&=& E_x\partial_{_{E_x}}f+ E_y\partial_{{E_y}}f\nonumber\\
&&+\nabla T_x \partial_{{\nabla T_x}}f+\nabla T_y\partial_{_{\nabla T_y}}f+\cdots\quad
\end{eqnarray}
by parameterizing $\bm{\mathcal{E}}$, ${\bf k}$, and ${\bf v}$ as $\bm{\mathcal{E}}=\mathcal{E}(\cos\theta, \sin\theta)$, ${\bf k}=k(\cos\phi, \sin\phi)$, and ${\bf v}({\bf k})=v(\phi)(\cos\xi(\phi), \sin\xi(\phi))$, respectively; we end up for nonequilibrium distribution function with,
\begin{eqnarray}
f(\theta, \alpha)-f_{0}&=&\left[A(\phi)\cos\theta+B(\phi)\sin\theta)\right]\mathcal{E}\nonumber\\
&&+\left[C(\phi)\cos\theta+D(\phi)\sin\theta\right]\nabla T
\label{eq9}
\end{eqnarray}
where, $A(\phi)=\partial_{E_{x}}f$, $B(\phi)=\partial_{E_{y}}f$, $C(\phi)=\partial_{\nabla T_{x}}f$, and $D(\phi)=\partial_{\nabla T_{y}}f$. By invoking the Eq. (\ref{eq9}) into  Eq. (\ref{eq4}) we obtain the following set of linear integral equations~\cite{Vyborny, zare2016, faridi}
\begin{eqnarray}
\cos\zeta(\phi)&=&\bar{w}(\phi)a(\phi)-\int d\phi'\frac{v(\phi^{\prime})}{v(\phi)}w(\phi, \phi^{\prime})a(\phi'),\label{ferd1}\\
\sin\zeta(\phi)&=&\bar{w}(\phi)b(\phi)-\int d\phi'\frac{v(\phi^{\prime})}{v(\phi)}w(\phi, \phi^{\prime})b(\phi').\qquad\label{ferd2}
\label{eq12}
\end{eqnarray}
with similar relations for $c(\phi)$ and $d(\phi)$. Here $w(\phi, \phi^{\prime})=(2\pi)^{-1}\int k^{\prime}dk^{\prime}w(k, k^{\prime})$ and $\bar{w}(\phi)=\int d\phi^{\prime}w(\phi, \phi^{\prime})$. Also, the quantities $A(\phi)=-ev(\phi)[-\partial_{\varepsilon}f_{0}]a(\phi)$, $B(\phi)=-ev(\phi)[-\partial_{\varepsilon}f_{0}]b(\phi)$, $C(\phi)=v(\phi)\left(\frac{\varepsilon-\mu}{T}\right)[-\partial_{\varepsilon}f_{0}]c(\phi)$ and $D(\phi)=v(\phi)\left(\frac{\varepsilon-\mu}{T}\right)[-\partial_{\varepsilon}f_{0}]d(\phi)$ are defined. Inserting solutions of Eqs. (\ref{ferd1}) and (\ref{ferd2}) into Eq. (\ref{eq9}) yields the exact solution of the Boltzmann equation up to the linear order in $\mathcal{E}$ and $\nabla T$ .
\par
By invoking the expression for $f(\theta, \phi)$ into Eq. (\ref{currents}) for the charge and heat currents, the response matrix, which relates the resulting generalized currents to the driving forces, can be expressed in terms of some kinetic coefficients $\mathcal{L}^{\alpha}$ as the following,
\begin{eqnarray}
\begin{pmatrix}
{\bf j} \\  {\bf j}^{q}
\end{pmatrix}=
\begin{pmatrix}
\mathcal{L}^{0} & -\mathcal{L}^{1}/eT\\
\mathcal{L}^{1}/e & -\mathcal{L}^{2}/e^{2}T
\end{pmatrix}
\begin{pmatrix}
\bm{\mathcal{E}} \\ -\nabla T
\end{pmatrix}
\label{eqbolt}
\end{eqnarray}
Diagonal response elements explain the electrical $\sigma$ and thermal $\mathcal{K}$ conductivities, and the two off-diagonal thermoelectric coefficients are related to each other through the Onsager relation. The Seebeck coefficient (thermopower) $\mathcal{S}=-\frac{1}{eT}(\mathcal{L}^{0})^{-1}\cdot\mathcal{L}^{1}$, describes the voltage generation due to the temperature gradient while Peltier coefficient $\Pi=T\mathcal{S}$ accounts for the heat current induction due to the charge current, respectively. The figure of merit, which is the ability of a material to efficiently produce thermoelectric power, is described by a dimensionless quantity denoted by $\mathcal{Z}T=\frac{\sigma\mathcal{S}^{2}}{\mathcal{K}}T$. All of the coefficients obey the relation
\begin{eqnarray}
\mathcal{L}^{\alpha}(\theta, \theta')=\int d\varepsilon\left[\frac{-\partial f_{0}}{\partial\varepsilon}\right](\varepsilon-\mu)^{\alpha}\sigma(\varepsilon; \theta, \theta')\label{cond_ene}
\end{eqnarray}
All of the thermoelectric properties described by $\mathcal{L}^{\alpha}$ can be found by calculating the generalized conductivity.

\end{document}